\documentclass[12pt]{iopart}

\usepackage{graphicx}

\begin{document}

\title[The $\beta-\gamma$ decay of $^{21}$Na]{The \boldmath $\beta-\gamma$ \unboldmath decay of \boldmath $^{21}$\unboldmath Na}

\author{N L Achouri$^1$, J C Ang\'elique$^1$, G Ban$^1$, B Bastin$^1$\footnote{Present address: GANIL, CEA/DSM-CNRS/IN2P3, 14076 Caen, France.}, B Blank$^3$, S Dean$^2$, P Dendooven$^2$, J Giovinazzo$^3$, S Gr\'evy$^4$, K Jungmann$^2$, B Laurent$^1$\footnote{Present address: CEA, DAM, DIF, F-91297 Arpajon, France.}, E Li\'enard$^1$, O Naviliat-Cuncic$^1$, N A Orr$^1$, A Rogachevskiy$^2$, M Sohani$^2$, E Traykov$^2$\footnote{Present address: Instituut voor Kern- en Stralingsfysica, K.U. Leuven, 3001 Leuven, Belgium.}, and H Wilschut$^2$}

\address{$^1$ LPC Caen, ENSICAEN, Universit\'e de Caen Basse-Normandie, CNRS/IN2P3, 14050 Caen, France.}
\address{$^2$ Kernfysisch Versneller Instituut, University of Groningen, 9747 AA-Groningen, The Netherlands.}
\address{$^3$ Centre d'\'Etudes Nucl\'eaires de Bordeaux Gradignan -
Universit\'e Bordeaux 1 - UMR~5797~CNRS/IN2P3, Chemin du Solarium, BP
120, 33175 Gradignan, France.}
\address{$^4$ GANIL, CEA/DSM-CNRS/IN2P3, 14076 Caen, France.}

\ead{achouri@lpccaen.in2p3.fr}

\begin{abstract}
A new and independent determination of the Gamow-Teller branching ratio in the $\beta$-decay of $^{21}$Na is reported. The value obtained of 5.13 $\pm$ 0.43 \% is in agreement with the currently adopted value and the most recent measurement. In contrast to previous experiments, the present method was based on the counting of the parent $^{21}$Na ions and the resulting 351~keV $\gamma$-rays without coincident $\beta$-particle detection. 
\end{abstract}


\submitto{\JPG}


\section{Introduction}
\label{intro}
\begin{figure}
\begin{center}
\resizebox{0.4\textwidth}{!}{%
\includegraphics{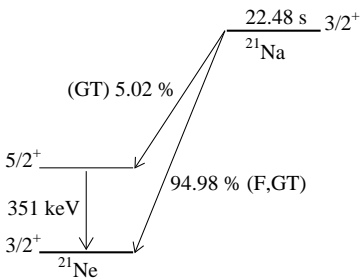}
}
\caption{$^{21}$Na decay scheme. The branching ratios adopted in the compilation~\cite{endt} are indicated. The extremely weak (4x$10^{-4}$ \%) decay branch to the $1/2^+$ state at 2794~keV is not shown.}
\label{fig:scheme}
\end{center}
\end{figure}

   The $\beta-\gamma$ decay of $^{21}$Na has attracted renewed interest recently~\cite{Iacob} owing to the role it plays in some experimental searches~\cite{scielzo, vetter} for physics beyond the Standard Model~\cite{herczeg}. In particular, the $\beta - \nu$ angular correlation in the decay of $^{21}$Na, which depends directly on the branching ratio of the decay to the $^{21}$Ne ground state, has been explored~\cite{scielzo, vetter}. As shown in figure~\ref{fig:scheme}, $^{21}$Na decays to the $^{21}$Ne ground state by a mixed Fermi (F) - Gamow-Teller (GT) transition and to the first excited state by a pure GT transition. Because of this simple level structure, a measurement of the GT branching ratio to the $^{21}$Ne first excited state enables the branching ratio to the $^{21}$Ne ground state to be deduced.

\begin{figure}
\begin{center}
\resizebox{0.5\textwidth}{!}{%
\includegraphics{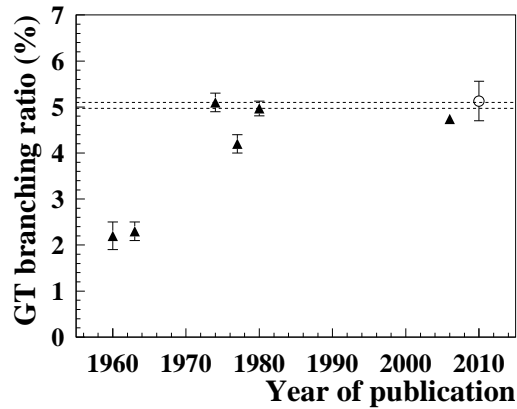}
}
\caption{Measurements of the $^{21}$Na $\beta-\gamma$ GT branching ratio. Those undertaken prior to the present work are represented by the triangles and are taken from~\cite{Iacob,talbert,arnell,alburger,azuelos,wilson}. The dashed lines represent the $\pm 1 \sigma$ limits of the adopted value~\cite{endt} (the weighted average of~\cite{alburger,wilson}) prior to\cite{Iacob} and the present measurement (open circle).}
\label{fig:br_an}
\end{center}
\end{figure}

 The measurements of the GT branching ratio made prior to the present work\footnote{We note that the present measurement was undertaken prior to the publication of the results of~\cite{Iacob}} are, as illustrated in figure~\ref{fig:br_an}, not all consistent. Importantly, these experiments were based on coincident detection of the 351~keV $\gamma$-ray with $\beta^+$-particles or with the 511~keV annihilation radiation. The goal of the present work was to determine the GT branching ratio with a different experimental method in order to provide an independent result. The principle of the method was to identify and count event-by-event $^{21}$Na nuclei implanted in a silicon detector telescope and to count the number of 351~keV $\gamma$-rays in singles mode --- without any $\beta^+$-coincidence --- using a high-purity germanium (HPGe) detector. Any uncertainties arising from the determination of the energy-dependent $\beta$-particle efficiency are thus avoided. As will be seen, the present method introduces other uncertainties which must be carefully evaluated.
 
 In order to validate the method and associated analysis procedures, the same measurement was performed for $^{22}$Mg for which the $\beta$-decay branching ratios are known with high precision~\cite{hardy}. In the present work, this measurement focused on the 582~keV $\gamma$-line which is the most intense in the $\beta$-decay of $^{22}$Mg.
 
 Finally, we note that preliminary results from an analysis of a subset of the data acquired here have been reported in~\cite{vetter,moi}.

\section{Experimental setup}
\label{sec:exp}

 The experiment was performed at the Kernfysisch Versneller Instituut (KVI) Groningen. The $^{21}$Na nuclei were produced by a (p,n) reaction with a 30 A MeV $^{21}$Ne primary beam on a hydrogen gas target located at the entrance of the TRI$\mu$P fragment separator~\cite{traykov1}. The $^{21}$Na nuclei represented 96\% of the secondary beam at the implantation point and were produced with an average rate of 150~pps.
  
\begin{figure}
\begin{center}
\resizebox{0.45\textwidth}{!}{%
\includegraphics{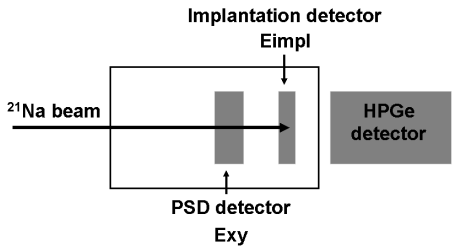}
}
\caption{Schematic drawing of the experimental setup located at the TRI$\mu$P separator final focus.}
\label{fig:setup}
\end{center}
\end{figure}

 The detection system located at the final focus of the separator is sketched in figure~\ref{fig:setup} and included a two-element silicon detector telescope,mounted in a vacuum chamber, for the ion identification. The first element, $E_{xy}$, was a Position Sensitive Detector (PSD), 300 $\mu$m thick, which provided an energy-loss measurement and a determination of the impact point of the incident ions. The second element, $E_{imp}$, was the implantation detector, 150 $\mu$m thick, which provided a measurement of the residual energy. A time-of-flight (ToF) was derived between the $E_{imp}$ detector timing signal and the cyclotron radiofrequency. A HPGe detector facing the chamber along the beam axis was used to measure the $\gamma$-rays which were detected in singles mode.
 
 For the $^{22}$Mg decay measurement, the beam was produced using the (p,2n) reaction with a 32 A MeV $^{23}$Na primary beam on the same hydrogen gas target. The $^{22}$Mg nuclei represented 65\% of the secondary beam at the implantation point with an average rate of 70~pps. 

\section{Analysis}

  The branching ratio of the $\beta$-decay of $^{21}$Na to the $^{21}$Ne first excited state, denoted $BR$, is given by:

\begin{equation}
BR = \frac{N_{\gamma}}{ \varepsilon_{\gamma} \; \times \; N_{imp} }
\label{branch}
\end{equation}

\noindent where $N_{\gamma}$ is the number of the 351~keV $\gamma$-rays detected (582~keV for $^{22}$Mg), $\varepsilon_{\gamma}$ is the efficiency of the HPGe and $N_{imp}$ the number of implanted ions. In the following, we will detail the determination of these quantities for each nucleus.

 Different sets of data taking runs were undertaken according to the various experimental parameters of importance. This enabled the possible sources of uncertainties associated with the data acquisition dead time, beam related backgrounds and the influence of the beam-spot size to be explored. For the two nuclei studied, the runs were grouped according to two criteria: the beam implantation mode and the slit settings at the final focus of the separator. Two implantation modes were employed: a beam-on/off mode whereby the beam was implanted for a fixed period and switched off while the decay activity was measured, and a continuous beam mode where the nuclei were implanted continuously and the activity measured simultaneously. To implant the secondary beam in the center of the silicon detector, the opening of the final focus slits --- located approximately 60 cm upstream of the implantation point --- was set to two different values. As some ions may have been implanted into the slits, a measurement with the beam impinging on the completely closed slits was performed in order to check if any $\gamma$-rays emitted from the decay of interest were detected by the HPGe detector.

\begin{figure} 
\begin{center}
\resizebox{0.55\textwidth}{!}{%
\includegraphics{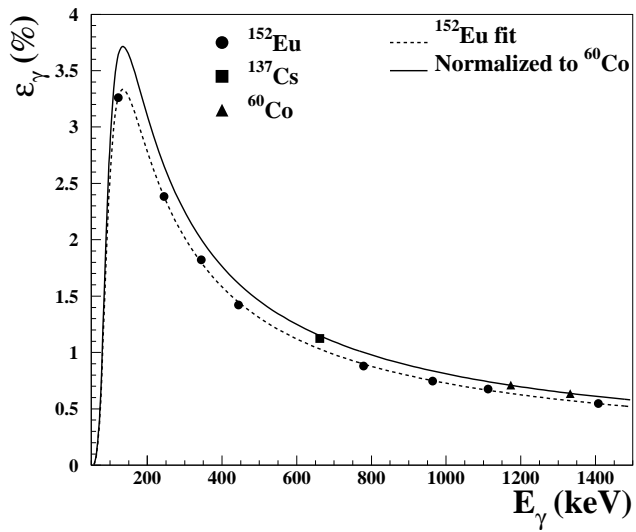}
}
\caption{Efficiency calibration of the HPGe detector obtained using $^{152}Eu$ (circles), $^{137}$Cs (square) and $^{60}Co$ (triangles) sources. The dashed line is the fit to the $^{152}Eu$ data and the full line is the normalization to $^{60}Co$ (see text).}
\label{fig:eff}
\end{center}
\end{figure}
 
\subsection{HPGe detector efficiency }
\label{effi}
  In order to establish the absolute efficiency of the HPGe detector, three sources were employed: $^{152}$Eu, $^{137}$Cs and $^{60}$Co. The absolute activities of the two last sources were known with precisions of 1\% and 0.1\%, respectively. The sources were placed at the implantation detector position so as to reproduce as precisely as possible the geometrical efficiency. In addition, measurements with the sources displaced by $\pm$ 1 cm from the beam axis were made in order to estimate the uncertainties in the efficiency arising from the finite size of the beam-spot. All of these measurements were in agreement within the error bars.

Owing to the wide range of lines in its spectrum, the $^{152}$Eu source was used to determine the shape of the efficiency curve (figure~\ref{fig:eff}). The fit to the $^{152}$Eu data was a function of the logarithm of energy polynomials~\cite{jackel}. The full line in figure~\ref{fig:eff} represents the normalization to the efficiency measured at 1173~keV and 1332~keV using the $^{60}$Co source, the absolute activity of which was known the most precisely. This efficiency has been corrected for the summing of the two $\gamma$-rays which are emitted in a cascade. The procedure to determine the summing correction is the same as that detailed for the $^{21}$Na analysis (section~\ref{Na21_gam}). The normalization to $^{60}$Co matched the efficiency measured at 662~keV with the somewhat less precisely known $^{137}$Cs source. Absolute efficiencies of $1.970 \pm 0.019 \%$ at 351~keV and $1.281 \pm 0.013 \%$ at 582~keV were thus deduced. The quoted errors include the uncertainties in the fit parameters, the normalization and the summing correction.

\begin{figure}
\begin{center}
\resizebox{0.6\textwidth}{!}{%
\includegraphics{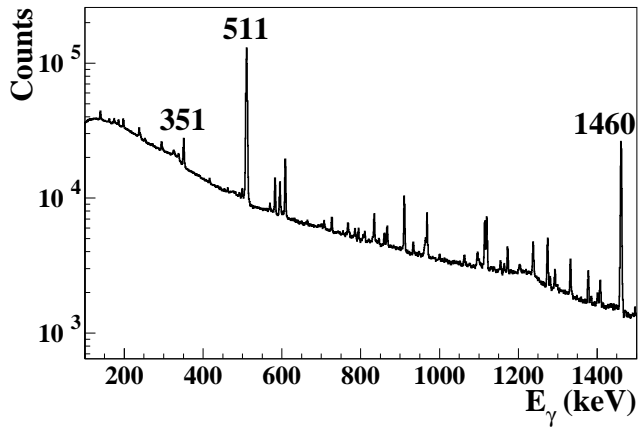}
}
\caption{$\gamma$-ray spectrum in the region of interest for data acquired with the $^{21}$Na beam. }
\label{fig:tout}
\end{center}
\end{figure}

\subsection{$^{21}$Na analysis}
\label{Na21}

\subsubsection{{\bf Determination of \boldmath $N_{\gamma}$ \unboldmath}}
\label{Na21_gam} 
\begin{figure}
\begin{center}
\resizebox{0.47\textwidth}{!}{%
\includegraphics{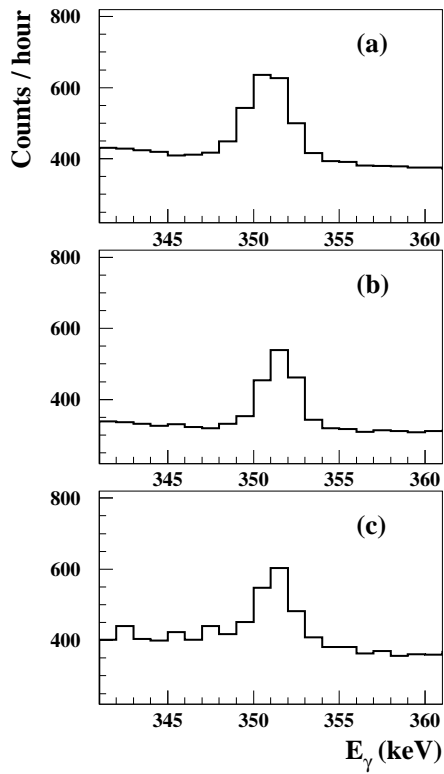}
}
\caption{The three figures show $\gamma$ spectra in the region of 351~keV: (a) $^{21}$Na implanted in the $E_{imp}$ detector, (b) a background spectrum (no beam) and (c) $^{21}$Na stopped on the fully closed final-focus slits. The number of counts was normalized with respect to the duration of each run.}
\label{fig:compar}
\end{center}
\end{figure}

Figure~\ref{fig:tout} shows a $\gamma$-ray spectrum obtained with the $^{21}$Na beam. The energy resolution (FWHM) was 2.2~keV at 344~keV and 2.6~keV at 1112~keV. The energies of the main $\gamma$-ray lines are indicated: 351 keV ($^{21}$Na decay), 511 keV (annihilation radiation) and a background peak at 1460 keV arising from $^{40}$K decay.
 Figure~\ref{fig:compar} shows a comparison between three spectra in the region of the transition of interest (351~keV) acquired under different conditions : \\ (a) with the $^{21}$Na ions implanted in the $E_{imp}$ detector; \\
(b) background spectrum (no beam); \\ 
(c) with the $^{21}$Na beam stopped on the fully closed final-focus slits.

 Figure~\ref{fig:compar}(b) exhibits a background peak at 352~keV arising from the decay of $^{214}$Pb, a member of the $^{238}$U decay chain~\cite{knoll}, while the peak in figure~\ref{fig:compar}(a) from the data acquired with the $^{21}$Na beam is slightly lower in energy and is broader. The width and the position of this peak (figure~\ref{fig:compar}(a)) are compatible with the presence of both the $^{21}$Na 351~keV transition and the 352~keV background line.
 
 The results for the run with the $^{21}$Na beam stopped on the final focus slits (figure~\ref{fig:compar}(c)) are compatible with the presence of only the 352~keV transition from $^{214}$Pb, indicating that any contribution during the data taking from ions implanted in them is negligible.

\begin{figure}
\begin{center}
\resizebox{0.45\textwidth}{!}{%
\includegraphics{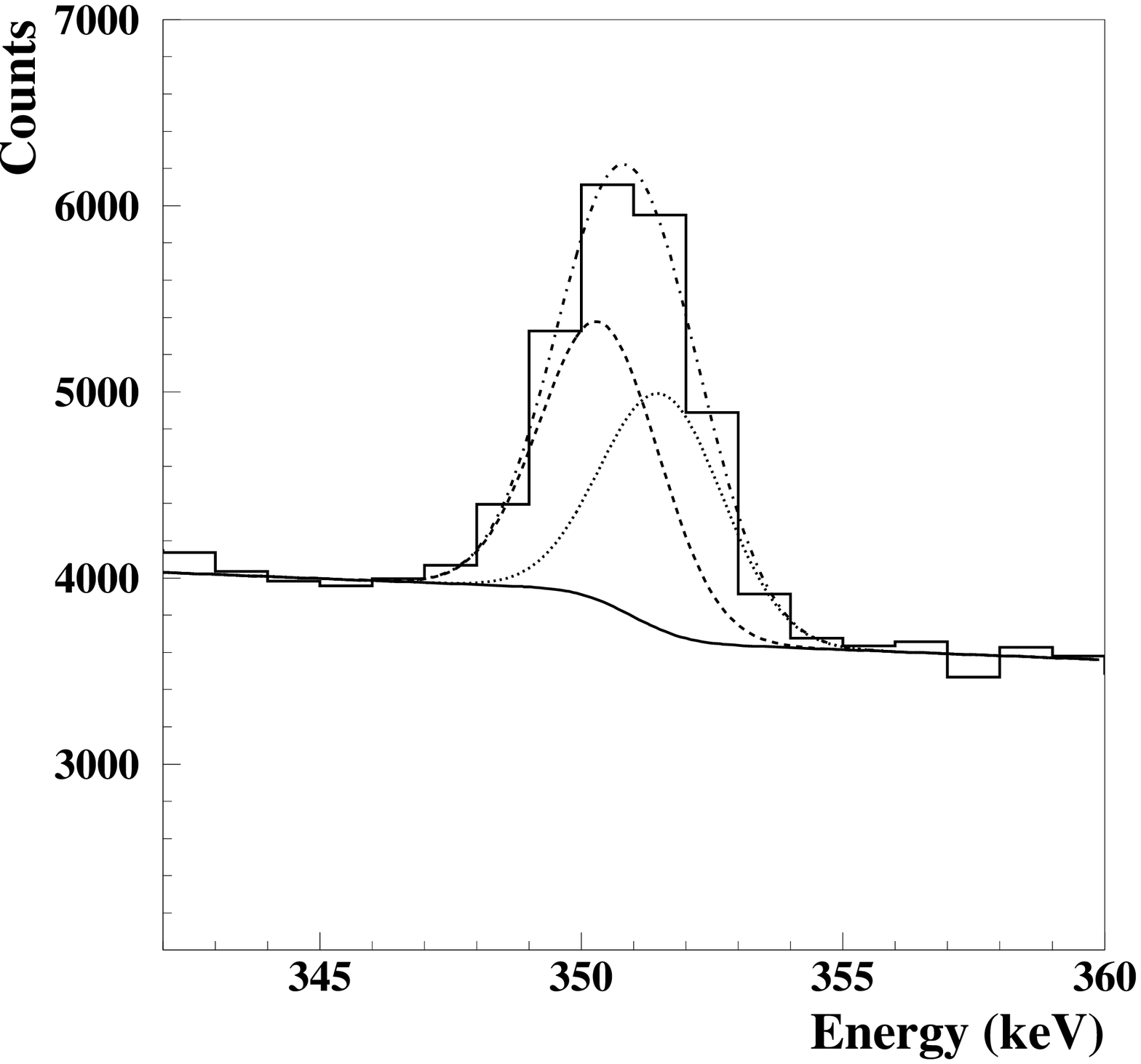}
}
\caption{Gamma-ray spectrum around 351~keV with the $^{21}$Na beam. The dashed line represents the contribution of the 351 $\gamma$-ray from the $^{21}$Na decay. The 352~keV background peak arising from $^{214}$Pb decay is shown as a dotted line. The background (full line) is a linear function combined with a step function. The sum of the three contributions is shown by the dashed-dotted line.}
\label{fig:fit}
\end{center}
\end{figure}

 To account for the $^{214}$Pb background peak, the intensity of the associated 352~keV $\gamma$-ray was normalized to that of the $^{40}$K line at 1460~keV in the background spectrum from the runs without beam. The corresponding ratio $R_0=23.4 \pm 0.4 \%$ remains the same in the spectrum with the $^{21}$Na beam. This ratio permitted the contribution of the $^{214}$Pb decay line to be fixed. A fit with a sum of two Gaussians and a linear function combined with a step function was used to describe the spectrum (figure~\ref{fig:fit}). The dashed line represents the contribution from the $^{21}$Na transition. The dotted line represents the contribution arising from the $^{214}$Pb background transition. Several constraints have been applied to the fitting of the spectrum in the region of interest: (1) the centroid of the background line was set to be 1.2~keV higher than that of the $^{21}$Na transition (the difference between the two energies of the transitions as tabulated in the literature~\cite{nndc}); (2) the widths of the two Gaussians were set equal and (3) the intensity of the background line was normalized, as described above, to the intensity of the $^{40}$K 1460~keV line in the same spectrum using the ratio $R_0$. As noted above, a linear function combined with a step function centered at the peak centroid was used to describe the underlying background (full line in figure~\ref{fig:fit}). This step function accounts for the $\gamma$-rays that Compton scatter in external material and the electrons which escape from the detector, as detailed by Helmer~{\it et al.}~\cite{helmer}. The fitting procedure, carried out using Minuit~\cite{minuit}, furnished $N_{\gamma}$ and the corresponding uncertainty. \\

\noindent {\bf Summing corrections} \\ 
  The summing of the $\gamma$-rays of interest with the 511 keV $\gamma$-ray can occur with the photopeak or with a Compton $\gamma$-ray. $N_{\gamma}$ should, therefore, be corrected for this effect to obtain the true number of emitted $\gamma$-rays. The following, should, therefore occur. \\
(a) When summed with the photopeak, the events of interest should lie at 862~keV. Unfortunately, a background peak arising from $^{208}$Tl ($^{232}$Th decay chain~\cite{knoll}) occurs at 860~keV. The same fitting procedure as described above was performed using the ratio between the $^{208}$Tl peak and the $^{40}$K peak determined from the background runs (without beam) to establish the contribution arising from the decay of $^{208}$Tl. The area of the sum peak so obtained was used to determine the total summing correction to $N_{\gamma}$. \\
(b) When summed with the 511 keV Compton $\gamma$-rays the correction is more complicated to obtain because the sum is distributed according to the Compton energy distribution.

Under these conditions, the total efficiency of the detector or the total-to-peak ratio (TTPR) has to be determined. When the TTPR is multiplied by the area of the sum peak ((a)) at 862~keV, one can determine the total amount of summing with the 511~keV events including the summing with Compton $\gamma$-rays. For this purpose, sources which emit a single $\gamma$-ray are required. The only one available in the present work was $^{137}$Cs, which with a single $\gamma$-ray at 662~keV permitted the total-to-peak ratio at this energy to be determined experimentally using the calibration run data. However, since the TTPR is needed at 511~keV, a detailed simulation of the HPGe detector and the surrounding material had to be undertaken using the MCNP package (version 4c)~\cite{mcnp} to determine the TTPR at 511 and 662~keV. At the latter energy the comparison of the measured TTPR with the simulated one shows that our simulation underestimates the TTPR (as found by Helmer et al.~\cite{helmer} and Venkataraman et al.~\cite{venka}), because the complete environment surrounding the detector (walls, for example) was not incorporated fully in detail in the simulation. As such, the simulated TTPR was normalized to the measured TTPR to account for these unavoidable deficiencies. Thus, the total-to-peak ratio at 511~keV was estimated to be $4.17 \pm 0.46$.

 The total correction to $N_{\gamma}$ corresponds to the area of the sum peak at 862~keV multiplied by this ratio to account for summings with both photoelectric and Compton $\gamma$-rays from the annihilation radiation.
The total summing correction applied to $N_{\gamma}$ was thus 7.51\%. In addition, the TTPR at 1252~keV which is the average energy of the two emitted $\gamma$-rays from the $^{60}$Co source was also measured using the calibration run data and determined using the simulation. This allowed the uncertainty of the approximations in the simulation of the geometry to be deduced. \\

\noindent {\bf Uncertainties} \\ 
  The uncertainties in $N_{\gamma}$, as listed in table~\ref{tab:error}, include: 
\begin{itemize}
\item the statistical uncertainty of 0.04\% in BR. 
\item the fitting procedure uncertainty of 0.17\% in BR.
\item the uncertainty arising from the choice of the form of the background. In order to estimate it, another fit was performed using only a linear function. The relative difference between the linear fit and the step function for all the runs was on average 0.4\% which leads to an uncertainty of 0.02\% in $BR$ (``Different fit functions'' in table~\ref{tab:error}).
\item the summing correction uncertainty which includes that of the area of the sum peak and that of the total-to-peak ratio. For the area of the sum peak, the uncertainty is relatively large owing to the small number of counts in the sum peak and to the uncertainty in the ratio between the $^{208}$Tl and the $^{40}$K lines ($3.74 \pm 0.25 \%$). For the total-to-peak ratio, in addition to the statistical uncertainty, an uncertainty of 11\%, which also accounts for the approximations in the simulation of the geometry, has been estimated as the difference between the simulated TTPR and that measured with the $^{60}Co$ source at the average energy of the two emitted $\gamma$-rays. The total summing correction uncertainty was thus estimated to be 0.37\% in $BR$.
\end{itemize}

\subsubsection{{\bf Determination of $N_{imp}$ }}
\label{Na21_imp} 

\begin{figure}
\begin{center}
\resizebox{0.5\textwidth}{!}{%
\includegraphics{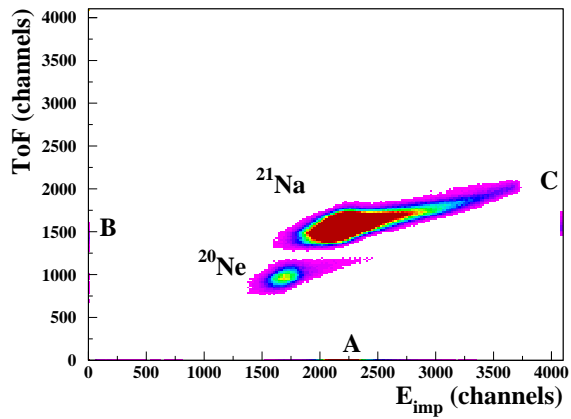}
}
\caption{ToF-residual energy spectrum derived from the implantation detector. See text for the description of groups A, B and C.}
\label{fig:impl}
\end{center}
\end{figure}

 The number of implanted ions was determined from the ToF and residual energy ($E_{imp}$) measurements as displayed in figure~\ref{fig:impl}. The main contaminant, the stable nucleus $^{20}$Ne, is well separated from the ions of interest. However, some events (A) located on the $E_{imp}$ axis have the correct residual energy for $^{21}$Na nuclei but no ToF. Other events (B) have the correct ToF but no $E_{imp}$ whilst some (C) have the correct ToF but $E_{imp}$ is in the overflow region. If the events comprising the three groups (A,B and C) correspond to implanted $^{21}$Na ions, then they should be included in the number of implanted $^{21}$Na ions ($N_{imp}$). \\

\noindent {\bf Corrections} \\
To determine the number of $^{21}$Na nuclei among the events denoted A, B and C, coincidences between them and $^{21}$Na ions identified in the $E_{xy}$-$E_{imp}$ spectrum have been searched for. For group A events, an average of 94\% were in coincidence with the $^{21}$Na ions in the $E_{xy}$-$E_{imp}$ spectrum. For group C events, an average of 96.5\% were in coincidence. For group B events, as $E_{imp}$=0, these events can not be in coincidence with $^{21}$Na in the $E_{xy}$-$E_{imp}$ spectrum. Given that it is impossible to ascertain if such events were actually stopped in the implantation detector, the number of group B events was considered as an uncertainty in the determination of $N_{imp}$.

 The correction to the number of implanted $^{21}$Na ions ($N_{imp}$) arising from the addition of the group A and C events was 1.16\%. \\

\noindent {\bf Uncertainties} \\
The uncertainties in $N_{imp}$, as listed in table~\ref{tab:error}, include:
\begin{itemize}
\item the statistical uncertainty which corresponds to 0.001\% in $BR$.
\item the uncertainty arising from drawing a different graphical contour --- on the same parameters --- used to select the $^{21}$Na events in figure~\ref{fig:impl}, which was estimated to be 0.002\% in $BR$.
\item the uncertainty arising from the addition of the A and C type events. This, when combined with the uncertainty attributed to group B events, represents an uncertainty of 0.006\% in $BR$.
\end{itemize}

\subsection{$^{22}$Mg analysis}
\label{Mg22}
 
\begin{figure}
\begin{center}
\resizebox{0.6\textwidth}{!}{%
\includegraphics{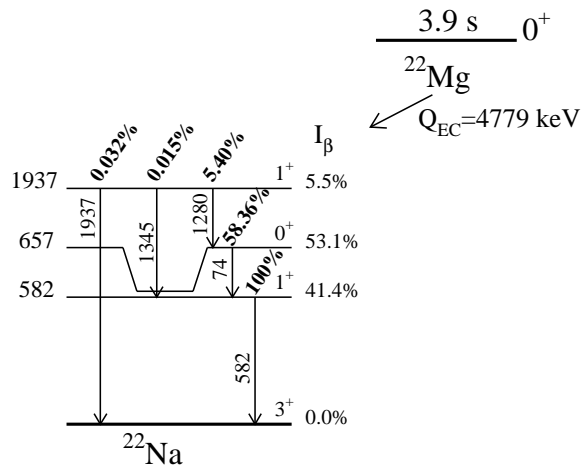}
}
\caption{$^{22}$Mg decay scheme. The relative intensities of the $\gamma$-transitions are from~\cite{hardy}.}
\label{fig:scheme_mg22}
\end{center}
\end{figure}

  As noted in the introduction, the $\beta$-decay of $^{22}$Mg, which is well known~\cite{hardy}, was investigated here in order to validate our experimental method and the various complications in the analysis.
  
   Figure~\ref{fig:scheme_mg22} shows the $^{22}$Mg decay scheme. The relative intensities of the $\gamma$-rays are those from~\cite{hardy}. As mentioned above, the present work concentrates only on the $\gamma$-ray line at 582~keV, which is the most intense. The ``branching'' ratio measured in this case is not the $\beta$-decay branching ratio from the ground state of $^{22}$Mg to one state in $^{22}$Na but --- as defined in equation~\ref{branch} --- is the ratio of the number of the $\gamma$-rays emitted at 582~keV relative to the number of decaying $^{22}$Mg nuclei. Indeed, the level at 582~keV is fed by higher-lying states populated in the $^{22}$Mg $\beta$-decay (figure~\ref{fig:scheme_mg22}). 

\begin{figure}
\begin{center}
\resizebox{0.6\textwidth}{!}{%
\includegraphics{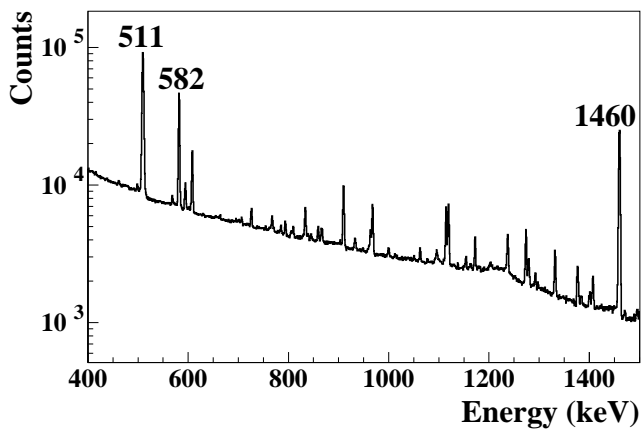}
}
\caption{$\gamma$-ray spectrum in the region of interest for data acquired with the $^{22}$Mg beam.}
\label{fig:gam_mg22}
\end{center}
\end{figure}

  Figure~\ref{fig:gam_mg22} shows the $\gamma$-ray spectrum in the region of interest for the $\beta$-decay of $^{22}$Mg. As seen in the analysis of the $^{21}$Na data, a background peak located at 583~keV arising from the $\beta$-decay of $^{208}$Tl is present and complicates the analysis.    
 The same fitting procedure used for the $^{21}$Na analysis has been applied, and in order to account for the background peak, the ratio $R_0$ between the $\gamma$-line at 583~keV from $^{208}$Tl decay and that at 1460~keV from $^{40}$K was determined from the background runs. Thus, $N_{\gamma}$ for $^{22}$Mg is given directly by the fit with the associated uncertainty listed in table~\ref{tab:error}.
      
 In this case also, the $\gamma$-ray line at 582~keV from the $^{22}$Mg decay can sum with the annihilation radiation at 511~keV resulting in a peak at 1093~keV. The number of counts of the sum peak was determined using the same procedure as in the $^{21}$Na analysis. To account for the summing with the 511~keV Compton $\gamma$-rays, this number was multiplied by the total-to-peak ratio at 511~keV as described earlier (section~\ref{Na21_gam}).  
 In addition, two other sums should be considered as a correction to $N_{\gamma}$, as a $\gamma$-ray at 74 keV is emitted in the $\beta$-decay of $^{22}$Mg~\cite{hardy}. In our case, the threshold on the HPGe was too high to detect this transition, but the summing with other $\gamma$-rays can be detected and can change the result in $N_{\gamma}$ for the $\gamma$-ray at 582~keV as follows :
\begin{enumerate}
\item  The sum peak at 585~keV arising from the 74 and 511~keV $\gamma$-rays cannot be disentangled from the $\gamma$-ray of interest. This contribution should be subtracted from $N_{\gamma}$ and was estimated as : \\
 $N_{74+511}= N_{imp} \times BR_{511} \times \varepsilon_{511} \times BR_{74} \times \varepsilon_{74}$.
\item The sum peak at 656~keV arising from the 582 and 74~keV $\gamma$-rays should be added to $N_\gamma$. This peak appears only in the $^{22}$Mg runs. A fit employing a Gaussian plus background was performed to obtain the number of summed $\gamma$-rays.
\end{enumerate}
 The correction applied to $N_{\gamma}$ was the sum of $N_{582+74}$ and $N_{582+511}$ corrected by the total-to-peak ratio at 511~keV from which $N_{74+511}$ was subtracted. It was found to be 3.58\% of $N_{\gamma}$ and the associated uncertainty was estimated to be 1.24\% in $BR$ (table~\ref{tab:error}). 

 The number, $N_{imp}$, of implanted $^{22}$Mg ions was determined as in the $^{21}$Na analysis (section~\ref{Na21_imp}) with the related uncertainties listed in table~\ref{tab:error}.
 
\section{Results and conclusions}

\begin{table}			
\caption{Uncertainties in the branching ratios for the decay of $^{21}$Na and $^{22}$Mg}
\label{tab:error}		
\begin{indented}
\lineup
\item[]\begin{tabular}{lcc}
\br	
 Uncertainty (\%)      	        &  $^{21}$Na    &  $^{22}$Mg  \\
\mr
 $N_{\gamma}$ statistics 				&  $ 0.04  $    &  $ 0.33 $  \\
 Fit of $\gamma$-peak	 					&  $ 0.17  $    &  $ 0.58 $  \\
 Different fit functions				&  $ 0.02  $    &  $ 0.03 $  \\
 Summing effects    	 					&  $ 0.37  $    &  $ 1.24 $  \\
 $N_{imp}$ statistics	 					&  $ 0.001 $    &  $ 0.04 $  \\
 $N_{imp}$ correction 	 				&  $ 0.006 $    &  $ 1.47 $  \\
 $N_{imp}$ contour 	 						&  $ 0.002 $    &  $ 0.04 $  \\
 $\gamma$-detection efficiency  &  $ 0.05  $    &  $ 1.01 $  \\
 $R_0$               	 					&  $ 0.11  $    &  $ 0.35 $  \\
\mr
 Total (\%)                     &  $ 0.43  $    &  $ 2.3  $  \\
\br
\end{tabular}
\end{indented}
\end{table}

  The branching ratio was determined for each set of runs made under different running conditions.
The branching ratios obtained from all the sets were in agreement within the uncertainties. The final quoted value is the weighted average. For each set, the uncertainty includes the uncertainty in $N_{\gamma}$ as described in section~\ref{Na21_gam} and the uncertainty in $N_{imp}$ as described in section~\ref{Na21_imp}. 

 The weighted average branching ratio was obtained using the ratio $R_0$ described in section~\ref{Na21_gam}. This ratio is subject to an uncertainty $\sigma_0$. The same analysis was performed for $R_0 \pm \sigma_0$. The difference between the resulting branching ratios was considered as an uncertainty added quadratically to the one obtained for the weighted average. In addition, the uncertainty in the $\gamma$-detection efficiency was also added quadratically to the uncertainty of the weighted average. The uncertainties described above are listed in table~\ref{tab:error} in terms of their impact on the branching ratio. 
 
\begin{table}			
\caption{Summary of measurements of the $^{21}$Na branching ratio}
\label{tab:branch}		
\begin{indented}
\lineup
\item[]\begin{tabular}{lc}
\br	
 Authors      	        					&  Branching ratio (\%)  \\
\mr
 Talbert et al.~\cite{talbert}		&  $2.2 \pm 0.3$   	   \\
 Arnell et al.~\cite{arnell}			&  $2.3 \pm 0.2$   	   \\
 Alburger et al.~\cite{alburger}	&  $5.1 \pm 0.2$   	   \\
 Azuelos et al.~\cite{azuelos}		&  $4.2 \pm 0.2$   	   \\
 Wilson et al.~\cite{wilson}			&  $4.97 \pm 0.16$ 	   \\
 Iacob el al.~\cite{Iacob} 				&  $4.74 \pm 0.04$ 	   \\
 Present work 										&  $5.13 \pm 0.43$ 	   \\
\br
\end{tabular}
\end{indented}
\end{table}

 The branching ratios deduced were $5.13 \pm 0.43 \%$ for $^{21}$Na and $101.3 \pm 2.3 \%$ for $^{22}$Mg. The branching ratio obtained for $^{22}$Mg is in agreement with the more precise value of $99.97\pm 0.19 \%$ of~\cite{hardy} \footnote{This value is obtained after normalization to 100\% of the relative $\gamma$-rays intensities of~\cite{hardy} at 583~keV and 1937~keV. Indeed, the sum of the intensities of these two $\gamma$-rays corresponds to the total $\beta$-decay strength of $^{22}$Mg.}. This result validates the experimental method and analysis techniques employed here.
 
 The present measurement for $^{21}$Na is in agreement with the previously adopted value of $5.03 \pm 0.13 \%$~\cite{endt} and with the recent result of Iacob et al. of $4.74 \pm 0.04 \%$~\cite{Iacob}. The poorer precision of the present result arises principally from the presence of the background $\gamma$-ray line at almost the same energy as the transition of interest and from the summing with 511~keV $\gamma$-rays (table~\ref{tab:error}). Table~\ref{tab:branch} summarizes all the measurements of the $^{21}$Na branching ratio in chronological order. 
 
\ack
We would like to thank the technical staff of KVI and LPC-Caen in the preparation and execution of the experiment. This work was supported by the European Commission within the Sixth Framework Programme through I3-EURONS (contract no. RII3-CT-2004-506065) and by the ``Stichting voor Fundamenteel Onderzoek der Materie'' (FOM program 48).

\end{document}